\documentclass{article}

\usepackage{PRIMEarxiv}

\usepackage[utf8]{inputenc} 
\usepackage[T1]{fontenc}    
\usepackage{hyperref}       
\usepackage{url}            
\usepackage{booktabs}       
\usepackage{amsfonts}       
\usepackage{nicefrac}       
\usepackage{microtype}      
\usepackage{lipsum}
\usepackage{fancyhdr}       
\usepackage{graphicx}       
\graphicspath{{media/}}     

\usepackage{amssymb}
\usepackage{amsmath}
\usepackage{amsfonts}
\usepackage{algpseudocode}
\usepackage{algorithm}
\usepackage{caption}
\usepackage{multirow}
\usepackage{graphicx}
\usepackage{textcomp}
\usepackage{xcolor}
\usepackage[caption=false]{subfig}
\usepackage{booktabs}

\title{Federated Learning Architecture: Data Privacy and System Security Approaches
}

\author{
  Cagdas Karatas \\
  Department of Computer Engineering\\
  Istanbul, Turkiye\\
  \texttt{cagdasfats@gmail.com} \\
   \And
   Hibanur Karadogan \\
  Department of Computer Engineering\\
  Istanbul, Turkiye\\
  \texttt{hkaradogan4455@gmail.com} \\
   \And
    Ahmet Yasin Ertug \\
  Department of Computer Engineering\\
  Istanbul, Turkiye\\
  \texttt{ahmetyasinertug@gmail.com} \\
     \And
     Busra Buyuktanir\\
  Department of Computer Engineering\\
  Istanbul, Turkiye\\
  \texttt{busra.buyuktanir@marmara.edu.tr} \\
     \And
     Kazim Yildiz \\
  Department of Computer Engineering\\
  Istanbul, Turkiye\\
  \texttt{kazim.yildiz@marmara.edu.tr} \\
  \And
  Gozde Karatas Baydogmus \\
  Loyola University Chicago, USA\\
  Biruni University, Turkiye \\
  \texttt{gkaratasbaydogmus@luc.edu} \\
}

\begin{document}
\maketitle

\begin{abstract}
This study explores the integration of homomorphic encryption and differential privacy techniques to enhance data privacy and security in Federated Learning (FL) systems. FL allows data to remain on local devices, eliminating the need for centralized data collection; however, sensitive information may still be leaked during model updates. To address this issue, homomorphic encryption enables computations on encrypted data, while differential privacy prevents the extraction of individual information through statistical techniques applied to model outputs. The proposed architecture was tested on the Framingham, Pima Indians Diabetes, and Bank Marketing datasets, revealing that enhanced privacy can be achieved without significantly compromising model accuracy. Furthermore, the impact of data heterogeneity among clients on model performance was analyzed, and it was concluded that strategies such as the careful selection of differential privacy parameters and training settings, along with the use of larger datasets, can improve the efficiency of FL. The findings demonstrate that privacy-preserving and high-performance artificial intelligence systems can be securely applied in sensitive domains such as healthcare and finance. 
\end{abstract}


\section{Introduction}

Today, artificial intelligence and machine learning are rapidly advancing with the development of models that require large amounts of data. However, concerns about data privacy and security lead to questioning centralized data collection methods \cite{kurtevaluating}. In this context, federated learning is an innovative approach that enables model training by processing data locally on devices without collecting it on a central server \cite{demir2026decentralized}. This method, first introduced by Google in 2016~\cite{konevcny2016federated}, offers significant advantages in both protecting individual users' privacy and training effective artificial intelligence models on large datasets.

Federated learning (FL)~\cite{buyuktanir2025federated} reduces privacy risks by ensuring that data remains on local devices; however, it presents some challenges in terms of security and privacy. Parameters transmitted during model updates can be analyzed by malicious actors, leading to the leakage of sensitive information. In particular, model inference and data reconstruction attacks can cause local data to be exposed. Additionally, the security of the central server used in FL processes is of critical importance; attacks on the server can compromise the security of the entire system. Therefore, it is necessary to integrate additional security measures such as homomorphic encryption and secure multi-party computation in FL applications~\cite{ozturk2025differential}.

Homomorphic encryption~\cite{acar2018survey,yilmaz2026homomorphic} allows storing data in an encrypted form while simultaneously enabling mathematical operations to be performed on this data. In other words, even if data is encrypted, certain mathematical operations can be performed directly on this encrypted data. This feature provides a great advantage, especially in situations that require analysis on sensitive data. For example, a health insurance company can perform analysis without directly accessing patients' medical records and create insurance offers through encrypted data.

This encryption method basically consists of three main steps~\cite{ogburn2013homomorphic}:
\begin{itemize}
    \item Encryption: Data is encrypted with a public key and shared with third parties.
    \item Operation: Predetermined mathematical operations are performed on encrypted data.
    \item Decryption: The operation result can only be decrypted with the private key owned by the authorized person and made meaningful.
\end{itemize}

Mathematically, homomorphic encryption is based on the following principle:

\begin{equation}
\label{eq1}
    E(a)\cdot E(b)=E(a*b)
\end{equation}

\begin{figure}[htbp]
    \centering
    \includegraphics[width=\linewidth]{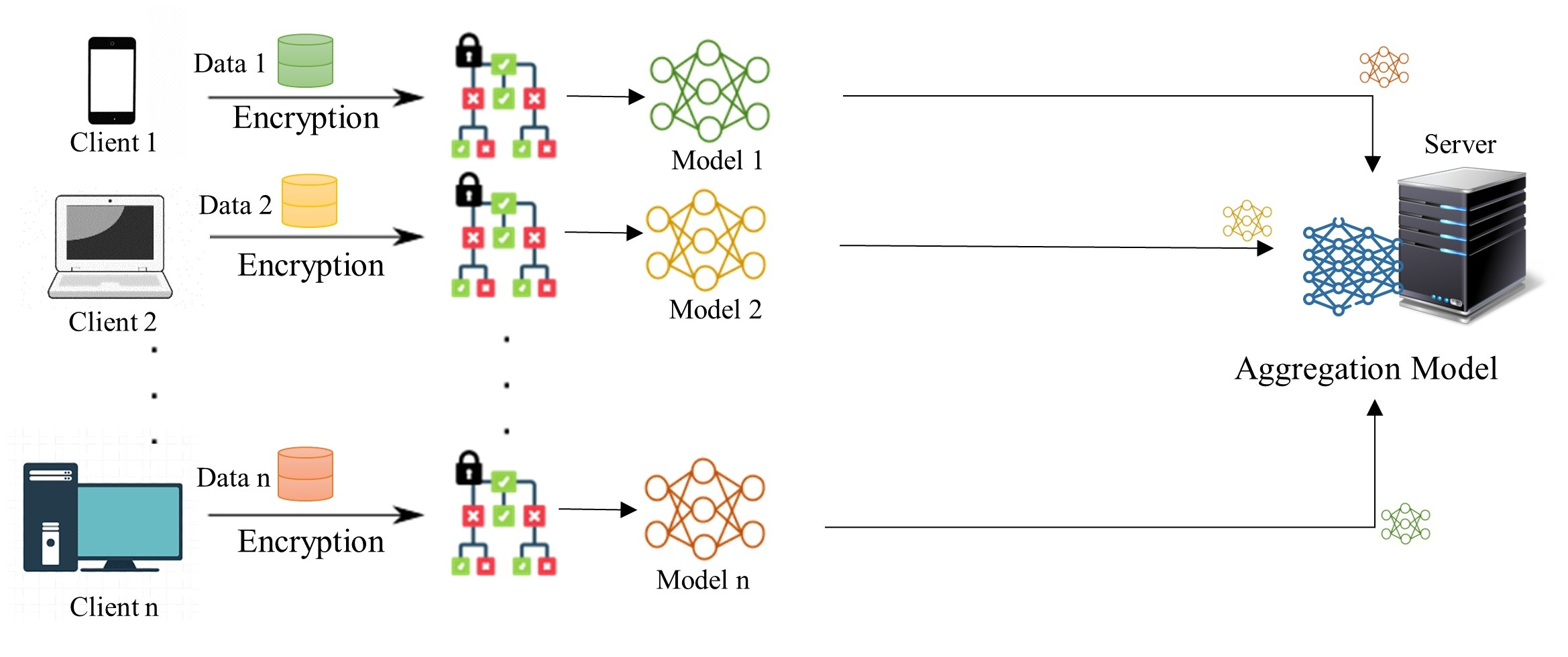}
    \caption{Privacy-preserving model training process in a federated learning architecture using homomorphic encryption. Diagram showing the federated learning architecture with homomorphic encryption, where clients perform local training on encrypted data and send model updates to a central server for aggregation.}
    \label{fl}
\end{figure}

In Equation~\ref{eq1}, $E(x)$ represents the encrypted form of the data, and ``*'' represents a specific mathematical operation (for example, addition or multiplication).
In Figure~\ref{fl}, data generated on clients is homomorphically encrypted and used for local model training. The trained models are securely transmitted to the server, where they are aggregated using federated learning algorithms~\cite{buyuktanir2023cba},\cite{buyuktanir2026enhancing}.

Homomorphic encryption systems are divided into different categories according to their operational capacity:
\begin{itemize}
    \item Partially Homomorphic Encryption: Allows only one type of operation such as addition or multiplication.
    \item Somewhat Homomorphic Encryption: Allows a certain number of addition and multiplication operations to be performed.
    \item Fully Homomorphic Encryption: Can perform both operations unlimited times.
\end{itemize}

Although fully homomorphic encryption is the most powerful model, its practical use is still limited due to high computational costs.

The combination of FL and homomorphic encryption contributes to creating secure artificial intelligence systems by increasing data privacy. Homomorphic encryption enables operations to be performed without decrypting data, making it possible to transmit model updates in an encrypted form during FL processes. In this way, even if model parameters are captured, their contents cannot be understood and data leakage is prevented.

However, sharing the same key among all participants in homomorphic encryption systems can lead to security risks. Therefore, the privacy of FL systems can be further strengthened with additional measures such as distributed key management, secure multi-party computation, and anonymization methods. The integration of these two technologies enables training artificial intelligence models without relying on central servers and secure sharing of model updates.

Within the scope of this study, a federated learning architecture strengthened with homomorphic encryption has been designed and the classification performance of this architecture on different datasets has been experimentally evaluated. Through tests conducted on Framingham, PID, and Bank datasets with different statistical characteristics, both the advantages provided by the system in terms of security and performance level have been analyzed~\cite{mahmood2014framingham,smith1988using}. Based on the metrics obtained as a result of the application, it has been observed that data heterogeneity between clients significantly affects model performance, therefore it has been concluded that strategies such as personalized model updates and adaptive weighted model aggregation can improve system performance. The findings reveal that privacy-preserving federated learning can provide secure and effective artificial intelligence solutions especially in areas containing sensitive data such as health and finance.

The main outline of this study can be summarized with the following points:

\begin{itemize}
    \item Identifying existing security and privacy vulnerabilities in FL processes.
    \item Analyzing the vulnerabilities of FL models against model inference and data reconstruction attacks.
    \item Evaluating the applicability and effectiveness of homomorphic encryption methods in FL environment.
    \item Proposing innovative and effective solutions to minimize security vulnerabilities in FL.
\end{itemize}

In this context, in the continuation of the study, existing studies in the literature on the subject will first be examined, then the details of the proposed method and the techniques used will be presented. Subsequently, experimental results will be evaluated to discuss the effectiveness of the method and the general conclusions of the study and improvements that can be made in the future will be addressed.

\section{Related Work}

A detailed literature review was conducted within the scope of this study. Table~\ref{tab:table1} provides general information about the studies reviewed in the literature.

\begin{table*}
  \caption{Literature review.}
  \label{tab:table1}
  \begin{tabular}{p{2.8cm}p{1.2cm}p{2.5cm}p{4.5cm}p{4.5cm}}
    \toprule
    Author(s) & Year & Method & Findings & Contribution \\
    \midrule
    \cite{li2020federated} & 2020 & Review & FL protects patient data privacy & Pioneered privacy-preserving healthcare AI \\
    \cite{xu2021federated} & 2021 & Experimental & FL achieved similar accuracy to centralized learning & Showcased FL's clinical applicability \\
    \cite{sen2024privacy} & 2023 & Experimental & Sensitive data can be reconstructed from FL updates & Highlighted critical privacy risks in FL \\
    \cite{mcmahan2017communication} & 2017 & Algorithmic & Efficient local training and reduced communication & Foundation of FL system architecture \\
    \cite{bonawitz2019towards} & 2019 & Implementation & Encrypted model updates for privacy & Applied FL at production scale in Android \\
    \cite{kairouz2021advances} & 2021 & Review & Comprehensive overview of FL technical issues & Guided future FL research directions \\
    \cite{hard2018federated} & 2018 & Application & On-device word prediction using FL & Proved FL's large-scale usability \\
    \cite{fang2021privacy} & 2021 & Experimental & Encrypted gradients with minimal accuracy loss & Efficient secure FL implementation \\
    \cite{lee2023hetal} & 2023 & Experimental & Transfer learning under encryption & Combined speed, privacy, and scalability \\
    \cite{rahulamathavan2023fhefl} & 2023 & System design & FL with fully homomorphic encryption & Defense against poisoning attacks \\
    \cite{laine2017simple} & 2017 & Tool & Open-source HE library & Real-world privacy-preserving ML \\
    \cite{gilad2016cryptonets} & 2016 & Experimental & Deep learning on encrypted data & Pioneered encrypted model inference \\
    \cite{masoud2019nonminimal} & 2019 & Experimental & Efficient learning with fewer parameters & Improved performance with less data \\
    \cite{dizaji2020training} & 2020 & Experimental & Fast adaptation with small data & Personalized healthcare model training \\
    \cite{panda2019certifying} & 2019 & Experimental & Efficient training with sparse labels & Improved accuracy in weak supervision \\
    \cite{wang2018joint} & 2018 & Experimental & Faster optimization in deep networks & Improved evolutionary learning efficiency \\
    \cite{wang2021statistically} & 2021 & Experimental & Increased model transparency and trust & Supports ethical and interpretable AI \\
    \cite{cheon2017homomorphic} & 2017 & Theoretical & Efficient computation on encrypted data & Enabled real-number ML on encrypted datasets \\
    \cite{canbay2020derin} & 2020 & Experimental & Heart disease prediction with $>$85\% accuracy on encrypted data & Practical privacy-preserving diagnosis \\
    \cite{fernandez2019aerogel} & 2019 & Review & Deep learning on encrypted data with $>$90\% accuracy & GDPR/HIPAA compliant HE-based ML \\
    \cite{applehe} & 2021 & Implementation & Encrypted computation in commercial products & Real-world use in Siri, iCloud, Health apps \\
    \bottomrule
  \end{tabular}
\end{table*}

In recent years, comprehensive research has been conducted on the security and privacy aspects of FL. Particularly in the healthcare sector, FL is used to train machine learning models while protecting patient data privacy. Additionally, studies on security threats and defense mechanisms are also being continued for FL to be applied fairly and reliably.

While FL offers significant advantages in protecting data privacy, it should be applied carefully due to the security and privacy risks it brings. Therefore, integrating appropriate privacy protection methods in the design and implementation of FL systems is of great importance.

\subsection*{Importance of FL in Healthcare in Terms of Data Privacy}

Today, in the healthcare sector, large amounts of personal and sensitive data are produced with the digitalization process, and the processing of this data enables the development of artificial intelligence-supported systems. However, protecting individual privacy in this process creates a critical problem in terms of ethical and legal concerns. In this context, FL stands out as a privacy-friendly machine learning approach by ensuring that patient data remains on local devices or hospitals without being centralized. Li et al. (2020) stated that federated learning significantly reduces privacy violations that may occur during the sharing of health data by ensuring that data remains at its source~\cite{li2020federated}. For example, multiple hospitals training a common model without physically sharing patient information both increases medical diagnosis accuracy and maximizes data security. In a study published in Nature Digital Medicine, it was stated that a model developed using federated learning with data collected from China and the USA for diagnosing COVID-19 cases gave results close to centralized learning performance~\cite{xu2021federated}. However, security vulnerabilities must also be taken into account to increase the applicability of this technology. In particular, attack types such as membership inference and model inversion carry the risk of indirectly exposing sensitive patient information from the trained model. In an illustrative example, Zhao et al. (2023) showed that information such as patients' age, gender, or disease type can be reconstructed from updated model parameters of federated learning systems~\cite{sen2024privacy}. Therefore, the integration of advanced privacy protection techniques such as differential privacy, homomorphic encryption, and secure multi-party computation is essential for federated learning to be safely applied in the healthcare field. All these findings reveal that federated learning is a paradigm that encourages collaboration in healthcare with privacy.

Developing effective artificial intelligence models while protecting customer data privacy in the finance sector is of great importance due to regulations and ethical responsibilities. At this point, federated learning enables different banks to collaborate without collecting their customer data in a central system. Especially in high-risk areas such as fraud detection, this approach increases the accuracy of fraud models and protects data privacy. In a study conducted jointly by Google Cloud and SWIFT, an advanced artificial intelligence system was developed using federated learning to prevent fraud in cross-border payments. This system enables financial institutions to create collective intelligence by collaborating without sharing user data~\cite{mcmahan2017communication}. Similarly, a system developed by Lucinity proposes the use of federated learning in combating financial crimes, enabling institutions to train crime detection models without sharing data belonging to their customers~\cite{truex2019hybrid}. Similar applications are also seen in academic literature. For example, in the study presented by Zhou et al., it was stated that federated learning-based credit scoring systems, especially for small and medium-sized banks, enable more accurate predictions by allowing customer data to be processed in a non-centralized manner~\cite{he2023credit}. One of the most important building blocks of federated learning is the Federated Averaging (FedAvg) algorithm developed by McMahan et al. in 2017. This study introduces a new machine learning paradigm that enables local model training on distributed devices (e.g., smartphones) without transferring data located on distributed devices to a central server. The authors propose a system design where user data privacy is protected while only model updates are sent to the central server. Especially to reduce communication costs, after several local training cycles are performed on devices, weights are sent to the server and a new model is created by averaging them here. This approach is groundbreaking in terms of communication efficiency and shows that it is possible to achieve high accuracy in scenarios where data privacy is ensured. The study is supported by both theoretical analyses and experiments conducted on MNIST and CIFAR-10 datasets. In this respect, it has both laid the conceptual foundation of federated learning and created a reference point for subsequent studies~\cite{mcmahan2017communication}. This comprehensive review study prepared by Kairouz et al. presents a holistic perspective on federated learning literature. The article systematically summarizes important developments in the federated learning field while discussing the fundamental problems faced by this technology and their solutions. The study addresses major technical challenges such as data heterogeneity (non-IID data distributions), hardware differences of client systems, irregularities in network connections, and privacy threats. Additionally, how privacy-enhancing techniques such as secure aggregation, differential privacy, and homomorphic encryption are applied in federated learning is also examined in detail. This study not only provides a technical analysis but also makes a significant contribution to the field in terms of indicating open problems suggested for future research. At this point where artificial intelligence and privacy intersect, it is extremely valuable in terms of shaping the research agenda~\cite{kairouz2021advances}. This study published by Li, Sahu, Talwalkar, and Smith addresses the specific challenges faced by federated learning in practice in a multifaceted manner. In particular, problems such as imbalance in data distributions, resource diversity between clients (e.g., processor power, battery life), and communication bandwidth limitations are defined as fundamental threats to the stability and efficiency of federated learning systems. Among the technical solutions proposed against these challenges in the article are various approaches such as asynchronous training, client sampling, model compression, and transfer learning. Additionally, the study discusses how federated learning is positioned in sectors where privacy is critical such as health, finance, education, and mobile devices and what policies should support it. In this respect, it is a valuable study that offers solutions both at the algorithmic level and at the systemic level, oriented towards practice~\cite{li2020federated}. This study published by Bonawitz et al. in 2019 details the technical architecture of the federated learning system developed by Google in the production environment. Especially in model training conducted with the participation of millions of users on Android devices, secure protocols to ensure data privacy are emphasized. The ``secure aggregation'' method introduced in the study allows model updates from clients to be processed by the central server without looking at their contents. This method enables collective learning while protecting the privacy of clients' individual contributions. Additionally, engineering details such as the system's resilience to network outages, client selection protocols, and how timing inconsistencies between devices are managed are also included. In this context, it is one of the most practical studies in the field in terms of showing how federated learning has been successfully transferred from theory to application~\cite{bonawitz2019towards}. The 2019 study by Yang et al. is an important resource that reveals the theoretical foundations and conceptual framework of federated learning. In the article, the difference between federated learning and classical centralized learning methods is clearly explained and what advantages these differences offer in terms of data privacy, regulatory compliance (e.g., GDPR), and user control are discussed. Additionally, the study presents potential applications of federated learning in various industrial areas such as finance, medicine, education, and transportation supported by example scenarios. The fact that some techniques that are difficult to understand at the conceptual level have been simplified with visuals and flowcharts makes this article very useful especially for researchers new to the subject~\cite{yang2019federated}. Hard et al., in this study where they introduce a concrete application of federated learning, examine the word prediction system in Google Gboard keyboard. In this system, data obtained from users' typing habits remain directly on the device, model training is done locally, and only learned model weights are sent to the central server. Thus, both user privacy is protected and the model can provide more accurate suggestions by adapting to personal habits. This study proves that federated learning is not only a theoretical concept but also a technology that can be successfully applied in large-scale systems. In this respect, the study presents an exemplary application in terms of artificial intelligence systems where privacy is prioritized~\cite{hard2018federated}. All these studies show that federated learning has the potential to provide not only privacy but also more reliable and explainable artificial intelligence solutions in the banking sector.

Homomorphic encryption offers an important step in protecting user privacy by allowing data to be processed in encrypted form. The system called FedML-HE aims to make model aggregation secure by integrating this encryption method into the federated learning framework. The system transmits only sensitive parameters encrypted during the training process, which significantly reduces both communication costs and computational load. FedML-HE has been tested especially on large-scale models such as ResNet-50 and BERT and has shown performance gains of approximately 10 to 40 times. This performance is remarkable in terms of both ensuring data privacy and achieving high-accuracy learning. Additionally, FedML-HE enables users to train high-accuracy models without the need for data sharing, promising hope for privacy-critical areas (such as health, finance).

Apple is one of the pioneering institutions in integrating homomorphic encryption into practical applications within the framework of its user privacy-oriented approach. Thanks to the company's open-source library called Swift-Homomorphic-Encryption, developers can benefit from cloud-based applications while keeping user data encrypted. This approach enables data to be processed without leaving the device, thus applications running on Apple devices both improve user experience and contribute to protecting personal data. Homomorphic encryption plays an important role, especially in services with high privacy sensitivity such as Siri, iCloud, and health applications. Apple's efforts in this field inspire the academic world in terms of demonstrating the applicability of homomorphic encryption in commercial products beyond the academic framework~\cite{applehe}. PFMLP (Privacy-Friendly Machine Learning Platform) is an original framework that shows how homomorphic encryption can be integrated with federated learning. This approach provides high-level data privacy without burdening the overall performance of the system by transmitting only important parameters encrypted. Gradients used in federated learning are secured with partial homomorphic encryption in PFMLP and can be processed on the server side without being decrypted. The study revealed that the model's accuracy reached up to 99\% in experiments conducted and showed only up to 1\% deviation compared to traditional methods. These results show that the system both provides high accuracy and offers a strong security mechanism~\cite{fang2021privacy}. HETAL (Homomorphically Encrypted Transfer Learning Algorithm) is a method that provides secure training by protecting client data privacy during transfer learning processes. This method encrypts client data with the CKKS homomorphic encryption scheme and sends it to the server, and model updates are performed directly in encrypted form. The system also makes the training process both faster and more effective with improvements such as validation-based early stopping mechanism and high-precision softmax function. In tests conducted on five different datasets, HETAL's total training time was less than one hour and the system provided a significant increase in security and efficiency compared to traditional transfer learning methods. In this respect, HETAL shows that homomorphic encryption offers advantages not only in terms of privacy but also in terms of speed and scalability~\cite{lee2023hetal}. FheFL (Fully Homomorphic Encryption-based Federated Learning) is an innovative system that maximizes the security of federated learning using fully homomorphic encryption. In this system, users send model updates they train locally to the server in encrypted form. These updates are then weighted according to participants' reliability scores and the central model is updated. In this way, model updates are both protected against malicious interventions and the integrity of the system is ensured. The secure aggregation mechanism proposed by FheFL provides effective defense against poisoning attacks frequently encountered in federated learning environments~\cite{rahulamathavan2023fhefl}. Microsoft SEAL (Simple Encrypted Arithmetic Library) is an open-source, C++-based library developed to integrate homomorphic encryption into practical applications. SEAL supports modern encryption schemes such as CKKS and BFV, making even very sensitive arithmetic operations possible to perform in encrypted form. This library is used in industrial applications as well as academic projects and stands out in areas where privacy is critical such as health, finance, and law. SEAL allows users to apply various machine learning algorithms without ever keeping their data in the open. For example, in areas requiring high privacy such as cancer diagnosis, data is kept encrypted and analyses are performed directly on this encrypted form. Microsoft's initiative reveals that homomorphic encryption is not only a theoretical concept but also a security layer that can be used in the real world~\cite{laine2017simple}. The interaction established between homomorphic encryption and machine learning algorithms creates a new paradigm in the analysis of privacy-sensitive data. Especially thanks to these methods that enable training data to be processed directly in encrypted form, results can be obtained without direct access to data and compliance with legal regulations (e.g., GDPR, HIPAA) can be ensured. The methods used in this interaction include adapted versions of Logistic Regression, SVM, and even some neural network architectures. For example, researchers have shown that simple deep learning structures can be trained with fully homomorphic encryption and accuracy rates can exceed 90\%. However, these operations usually require high computational costs, which slows down the widespread use of this technology. Nevertheless, thanks to hardware accelerators and optimization techniques, this interaction is gradually becoming more applicable~\cite{fernandez2019aerogel}. The groundbreaking study called Cryptonets is one of the first projects to show that deep learning can be performed on encrypted data. In this study, researchers trained a neural network model on the MNIST dataset with completely encrypted inputs and achieved successful results with only a small loss in accuracy rate. This approach enables model training on medical or financial data that cannot be directly accessed due to data privacy. Cryptonets basically uses the BFV homomorphic encryption scheme and performs basic mathematical operations (addition, multiplication) predominantly in encrypted form. One of the biggest advantages offered by this system is that accurate predictions can be made without decryption; this both protects data owners' privacy and allows service providers to perform analysis more securely~\cite{gilad2016cryptonets}. The healthcare sector is one of the areas where personal data is processed in the most sensitive form, so homomorphic encryption plays a major role here. Especially non-centralized and high-privacy information such as hospital data can be analyzed thanks to this technology, and applications such as disease diagnosis and treatment recommendations can be performed on encrypted data. One of the studies trained a machine learning model developed for heart disease diagnosis under homomorphic encryption and achieved a success rate of over 85\%. Another noteworthy element here is that collective analysis can be performed without data owners sharing their data, that is, personal privacy is protected throughout the entire process. These types of applications have become even more important with the spread of remote health services, especially during the pandemic period~\cite{canbay2020derin}. The CKKS (Cheon-Kim-Kim-Song) homomorphic encryption scheme is a method that stands out especially in approximate calculations of floating-point data. This scheme is widely preferred because it is more suitable for real number calculations naturally used in machine learning applications. CKKS enables users to perform neural network calculations both securely and practically. Additionally, operations performed with CKKS are more efficient and have lower latency compared to other fully homomorphic systems. In some studies conducted, it has been observed that CKKS-based deep learning algorithms give results with only 3\% accuracy loss when working on encrypted data, but the privacy gain is much higher. This scheme is particularly preferred in areas such as financial analysis, biomedical research, and personalized recommendation systems~\cite{cheon2017homomorphic}.

In conclusion, homomorphic encryption plays a groundbreaking role in ensuring the security of artificial intelligence-based technologies such as machine learning and federated learning, especially in today's information age where data privacy is of great importance. This encryption method both protects user privacy and offers solutions compliant with regulatory legislation (such as GDPR, HIPAA) by allowing data to be processed in encrypted form. Systems such as FedML-HE, PFMLP, HETAL, and FheFL have shown that models providing privacy protection with high accuracy and efficiency can be realized through the integration of homomorphic encryption into federated learning. The practical applications of technology giants such as Apple and Microsoft in this field reveal that homomorphic encryption can create a reliable security layer in commercial products beyond the academic framework. Especially systems developed with modern schemes such as CKKS and BFV enable neural networks and other machine learning algorithms to be trained on encrypted data. Additionally, pioneering studies such as Cryptonets have proven that this technology can give high-accuracy results even with limited resources. Applications in areas where privacy is critical such as health and finance enable analysis to be performed without users sharing their data, both protecting individual privacy and enabling information production for the benefit of society. The results obtained in comprehensive tests show that homomorphic encryption is not only a security measure but also a scalable, efficient, and ethically sensitive technology. In this context, homomorphic encryption is a candidate to be a fundamental building block in the future of artificial intelligence and data science applications.

Differential learning is generally a learning approach used in the field of machine learning and artificial intelligence to increase the accuracy of the model. This approach aims to offer a more efficient method during parameter updates and model training. Although differential learning is usually associated with gradient-based methods, it includes special strategies that allow updating only important parameters of the model. This enables the model to train with less data, allowing faster results to be obtained in the learning process. In a study conducted in 2018, it was observed that differential learning significantly increased accuracy in complex models such as deep neural networks. Researchers have shortened training time and achieved high accuracy rates by updating model parameters with smaller and more meaningful steps. It has been argued that this method gives much more efficient results in large datasets as well.

Differential learning is effectively used especially in image recognition and natural language processing (NLP) areas. While deep learning models used in these areas usually require large datasets, differential learning stands out as a method that enables efficient training of such models. In a study conducted in the field of image recognition, it has been shown that the differential learning approach increases efficiency by considering only the most important features during the model's training process. Similarly, in the NLP field, in tasks such as language modeling and text classification, model parameters have been updated more carefully with differential learning techniques, and higher accuracy has been achieved with less data. This provides a significant advantage, especially in training large language models. The successes in this field prove that differential learning is an efficient and flexible method~\cite{masoud2019nonminimal}. When differential learning is combined with meta-learning, it can make the training of personalized models faster and more efficient. Meta-learning can be defined as ``learning to learn'' and this technique aims to develop a model's ability to quickly adapt to new tasks. Differential learning, in the context of meta-learning, enables the model to be effective in new tasks by optimizing the initial parameters of the model more quickly. A research revealed that differential learning, especially in models trained on personal datasets, provides faster and more effective results in areas such as personalized health applications by better managing data heterogeneity and adaptation. When used together with meta-learning, differential learning accelerates the learning process, enabling an effective model to be created with less data~\cite{dizaji2020training}. Weakly supervised learning aims to optimize learning processes in situations where labeled data is insufficient. Differential learning makes significant contributions to weakly supervised learning systems. Because differential learning enables model parameters to be updated correctly with less information. This enables the model to learn more efficiently when labeled data is missing. Especially in areas where difficult and costly processes such as labeling medical images exist, significant successes have been achieved with improvements obtained through differential learning. In a study conducted, it was observed that the differential learning method increased accuracy rates especially in tasks such as image classification and anomaly detection in a weakly supervised environment. This method offers an efficient solution in many applications based on learning with weak labeling or unlabeled data~\cite{panda2019certifying}. Differential learning is used in developing evolutionary learning strategies by combining with genetic algorithms. While genetic algorithms try to optimize solution candidates in a population through an evolutionary process, differential learning makes parameter updates more efficient in this process. Such a combination provides a significant advantage especially in complex optimization problems. The integration of genetic algorithms with differential learning method has accelerated the optimization process in deep learning networks and more effective results have been obtained. In a study, it was proven that the combination of evolutionary algorithms with differential learning provides more accurate solutions with fewer iterations and provides more efficient learning. This shows great development in applying genetic algorithms to more complex problems~\cite{wang2018joint}. Differential learning can also be used as a technique to increase the transparency of the model. Making the decision-making processes of models more understandable is usually an important obstacle that prevents the widespread use of complex models. Differential learning makes the learning processes of the model more transparent by making changes only on meaningful and important parameters of the model. This is especially important for ethical and explainable artificial intelligence applications. Especially in decision support systems and critical areas such as health, differential learning methods are used to provide more transparency about how the model makes decisions based on which data. In a study conducted, it was seen that models trained with differential learning are more explainable and transparent compared to other methods. This increases the reliability of the model and enables it to be accepted by a wider user base~\cite{wang2021statistically}.

As a result of these studies, the differential learning approach provides significant contributions in terms of both efficiency and flexibility in the fields of machine learning and artificial intelligence. Especially this method, which focuses on updating only meaningful parts of model parameters, makes it possible to obtain higher accuracy results with less data. In data-intensive areas such as image recognition and natural language processing, differential learning enables models to be trained faster and more efficiently. When combined with meta-learning, it enables personalized models to be trained more effectively with fewer examples, while in weakly supervised learning environments, it helps achieve high accuracy rates even when labeled data is scarce. Thanks to its integration with genetic algorithms, evolutionary learning processes can be optimized in shorter time and more efficiently. In the context of reinforcement learning, it has accelerated the learning process by enabling agents to establish more effective interactions with the environment and make better policy updates. Additionally, differential learning contributes to the development of transparent and explainable artificial intelligence applications by increasing the interpretability of models. This is of great importance especially in terms of increasing model reliability in health and ethical decision support systems. In general, differential learning stands out as an innovative and effective learning strategy that offers multi-dimensional advantages to machine learning systems in terms of speed, accuracy, data efficiency, personalization, interpretability, and low data requirement.

\section{Methodology}

\subsection{Dataset}

\subsubsection{Framingham Dataset}
In this study, cardiovascular disease prediction was performed using individuals' sensitive health data in the FL environment, and advanced data protection techniques were applied to increase privacy and security in this process~\cite{mahmood2014framingham}. The dataset used is an open dataset derived from the Framingham Heart Study and accessed through the Kaggle platform~\cite{framingham2020kaggle}. In this dataset containing 15 variables that directly affect public health such as age, gender, blood pressure, cholesterol levels, diabetes status, smoking, body mass index (BMI), heart rate from a total of 4,240 individuals, the target variable was defined as a binary classification label (TenYearCHD) indicating whether the individual will have coronary heart disease in the next 10 years. The Framingham dataset was preferred due to its low missing data rate, structural integrity, and clinically meaningful health variables. This dataset with a community-based sampling structure creates a suitable test ground for FL scenarios because it consists of parameters that can be measured by individuals or health institutions, and supports on-device processing, which is one of the basic principles of FL, by allowing data to be processed on local devices without being transferred to a central server.

In the dataset preprocessing process, some columns that were evaluated as not directly related to the target variable (such as education level) or containing high rates of missing data were removed; remaining missing observations were cleaned. Then, input variables were normalized with StandardScaler to prevent the model from being affected by different scales between features. To prevent class imbalance from negatively affecting model performance, samples belonging to the minority class were increased using the SMOTE (Synthetic Minority Over-sampling Technique) method to achieve a balanced data distribution. These steps formed a critical foundation for developing prediction models with high accuracy and generalizability in the FL environment. The dataset was randomly split into 90\% training and 10\% testing to evaluate the model's generalization ability.

\subsubsection{Pima Indians Diabetes Dataset}
In this study, the second dataset used for diabetes prediction is the Pima Indians Diabetes Database, which contains clinical data of Pima Indian women aged 21 and over living in Arizona, USA~\cite{smith1988using}. This dataset of 768 individuals contains eight features: number of pregnancies, plasma glucose concentration, diastolic blood pressure, triceps skin thickness, 2-hour serum insulin level, body mass index (BMI), diabetes pedigree function, and age. The target variable is defined as a binary classification label (Outcome) indicating whether the individual is diabetic or not. The dataset is structurally consistent except for cases where missing values are coded as zero and contains clinically meaningful health variables. With these characteristics, it provides a suitable test ground for FL scenarios in terms of processing data on local devices and preserving privacy.

In the data preprocessing stage, zero values observed in Glucose, BloodPressure, SkinThickness, Insulin, and BMI variables were evaluated as biologically invalid and these values were recoded as missing data (NaN). Then, missing values were filled with the median of each column to ensure data integrity. Outliers in numerical columns were limited with lower and upper bounds using the interquartile range (IQR) method, thus reducing the possible effects of extreme values on model performance. Following these processes, input data were normalized with the StandardScaler method, then samples belonging to the minority class were increased using the SMOTE (Synthetic Minority Over-sampling Technique) algorithm to achieve a balanced class distribution to eliminate class imbalance. This systematic preprocessing process formed a solid foundation for developing reliable and generalizable prediction models in the FL environment. The dataset was randomly split into 90\% training and 10\% testing to evaluate the model's generalization ability.

\subsubsection{Bank Marketing Dataset}
This study also examined the Bank Marketing dataset from the UCI Machine Learning Repository. This dataset is based on real customer data related to direct marketing campaigns of a bank in Portugal. This dataset consisting of 41,188 observations and 17 attributes contains the success status of campaign calls made to customers via telephone. The dataset includes both qualitative and quantitative variables related to demographic and campaign interaction such as age, education level, marital status, income status, previous campaign contacts, and previous relationships with the bank. The target variable is a binary classification label indicating whether a customer subscribed to a term deposit product as a result of the marketing campaign (yes/no).

The dataset is widely used for modeling customer behaviors in the banking sector and particularly constitutes a suitable example for testing privacy-preserving machine learning approaches. In this aspect, in the context of Federated Learning (FL) architecture, it realistically represents the scenario of processing customer data belonging to different branches on local devices without transferring to a central server. In areas requiring high privacy, such as banking, FL architecture allows both individual data privacy to be preserved and a contribution to the central model. For this reason, the Bank Marketing dataset was preferred in this study to evaluate the application potential of FL.

Within the scope of the data preprocessing process, first, outliers in numerical variables were detected using the interquartile range (IQR) method, and values outside the lower and upper bounds were clipped to extreme points. This method was applied to prevent excessive extreme values from disrupting the model's learning process. Then, normalization was performed with the StandardScaler method to prevent input variables with different scales from negatively affecting model performance. Categorical variables were converted to numerical form using the One-Hot Encoding technique so that machine learning algorithms could work effectively with this data.

The significant imbalance observed between the classes of the target variable in the dataset can weaken the model's prediction ability for the minority class. To address this problem, downsampling was applied by randomly sampling from observations belonging to the majority class, achieving a balanced distribution between classes. In this way, the model was able to learn with equal sensitivity to both classes and more reliable classification results were obtained.

All these systematic data preprocessing steps aimed to increase both the generalizability and statistical consistency of the model developed in the FL environment, resulting in a more robust prediction infrastructure.

Table~\ref{tab:data-preprocessing} presents a comparison of data preprocessing steps applied to the Framingham, PID, and Bank datasets.

\begin{table}
\centering
  \caption{Comparison of data preprocessing procedures.}
  \label{tab:data-preprocessing}
  \begin{tabular}{lccc}
    \toprule
    Feature & Framingham & PID & Bank \\
    \midrule
    Missing data handling & $\checkmark$ & $\checkmark$ & $\times$ \\
    Outlier handling      & $\checkmark$ & $\checkmark$ & $\times$ \\
    Standardization       & $\checkmark$ & $\checkmark$ & $\times$ \\
    One-hot encoding      & $\checkmark$ & $\checkmark$ & $\times$ \\
    SMOTE                 & $\checkmark$ & $\checkmark$ & $\times$ \\
    \bottomrule
  \end{tabular}
\end{table}

\subsection{Methods}

\subsubsection{CKKS Homomorphic Encryption}
In this study, homomorphic encryption techniques were utilized to develop machine learning models while preserving the privacy of sensitive health data. Particularly, the CKKS (Cheon--Kim--Kim--Song) algorithm, which allows processing of decimal numbers in encrypted form and has high computational efficiency, was preferred. The CKKS algorithm is an approximate computational homomorphic encryption algorithm based on BGV encryption. CKKS reduces communication overhead by integrating batch encryption and increases computational efficiency by allowing small errors. Additionally, CKKS supports operations such as addition, subtraction, multiplication, and division while maintaining cyclic security. Furthermore, it enables encryption and computation of decimal numbers.

The CKKS algorithm is fundamentally based on the BGV (Brakerski--Gentry--Vaikuntanathan) structure, but differently, it consciously tolerates accuracy loss (approximation error) during encryption and decryption.

Encryption is performed in polynomial rings:
\begin{equation}
\label{eq2}
R = \mathbb{Z}[X]/(X^N + 1) \quad \cite{cheon2017homomorphic}
\end{equation}

In Equation~\ref{eq2}, $N$ is the security parameter of the system, being an even number~\cite{cheon2017homomorphic}. The encryption process begins with converting real numbers encoded at constant precision into polynomials by embedding them into this ring structure. In this process, data is first embedded in the complex plane $\mathbb{C}^n$, then numbers are encoded with quantization as in Equation~\ref{eq3}.

\begin{equation}
\label{eq3}
\text{Encode}(z) = \left\lfloor \Delta \cdot z \right\rfloor
\end{equation}

The encryption process can be summarized as Equation~\ref{eq4}:
\begin{equation}
\label{eq4}
\text{Enc}(m) = (c_0, c_1) = (m + \Delta \cdot e,\; a)
\end{equation}

The decryption process is performed as Equation~\ref{eq5}:
\begin{equation}
\label{eq5}
m' = \left\lfloor \frac{c_0 + c_1 \cdot s}{\Delta} \right\rceil
\end{equation}

In Equation~\ref{eq5}, $s$ is the secret key, and approximately the original value of $m$ is recovered as a result of the operation.

Equations~\ref{eq6} and~\ref{eq7} illustrate the additive and multiplicative properties of homomorphic encryption~\cite{halevi2014algorithms}. The CKKS algorithm enables operations such as addition and multiplication on encrypted values to yield results equivalent to those performed on plaintext values before encryption:

\begin{equation}
\label{eq6}
\text{Enc}(m_1) + \text{Enc}(m_2) \Rightarrow \text{Enc}(m_1 + m_2) \quad \cite{halevi2014algorithms}
\end{equation}

\begin{equation}
\label{eq7}
\text{Enc}(m_1) \cdot \text{Enc}(m_2) \Rightarrow \text{Enc}(m_1 \cdot m_2) \quad \cite{halevi2014algorithms}
\end{equation}

This capability of the CKKS scheme is particularly advantageous in applications involving sensitive health data, as it allows meaningful computations to be performed on encrypted data while maintaining data privacy. However, a certain amount of noise ($\epsilon$) is introduced during these operations, which is an inherent characteristic of the approximate computation nature of CKKS. Controlling the level of this error is crucial to ensure the accuracy of the decrypted results.

The amount of error $\epsilon$ that occurs during these operations must be kept under control. After each operation, the error increases somewhat; this is a natural consequence of CKKS's approximate computation feature.

In conclusion, the reason we preferred the CKKS algorithm for homomorphic encryption in our project is that it can provide data security and computational efficiency simultaneously. CKKS allows efficient computations to be performed while increasing data security by offering small error tolerance when operating on encrypted data. This provides great advantages, especially in projects where sensitive health data is processed.

\subsubsection{Differential Privacy}
Today, the protection of individual data emerges as one of the most critical problems along with advances in the field of data analytics. In this context, Differential Privacy (DP) is an effective approach that mathematically limits the effect of individuals' contribution to data on analysis results and thus strongly guarantees privacy. This study addresses the basic definition of differential privacy, its mathematical foundations, applied mechanisms, and its uses in various practical areas~\cite{dwork2006calibrating}.

Modern data analysis systems process large volumes of data in a wide range of areas such as health, economics, and social sciences. However, protecting individuals' privacy is one of the most fundamental ethical responsibilities that such systems must meet. Traditional data anonymization methods may be insufficient as they cannot provide complete assurance against the re-identification of individuals. Differential privacy, developed to address this inadequacy, provides a robust framework aimed at safely hiding the presence or absence of individuals in a specific dataset through statistical outputs.

For an algorithm to provide differential privacy, the probability distributions of its outputs on two ``neighboring'' datasets (sets that differ only in terms of one individual's data) must be close to each other. A randomized algorithm $A$ is differentially private if for any two neighboring datasets $D$ and $D'$ and any subset $S$ of $A$'s output set satisfies the inequality. In Equation~\ref{eq9}, $\varepsilon$ is the privacy parameter, and as its value decreases, the privacy guarantee obtained increases~\cite{dwork2014algorithmic}.

\begin{equation}
\label{eq9}
\Pr[A(D) \in S] \leq e^{\varepsilon} \cdot \Pr[A(D') \in S] \quad \cite{dwork2014algorithmic}
\end{equation}

In some scenarios, the strict requirements of differential privacy can be relaxed. In these cases, $(\varepsilon, \delta)$-differential privacy definition is used, which allows for a small error probability $\delta$:
\begin{equation}
\label{eq10}
\Pr[A(D) \in S] \leq e^{\varepsilon} \cdot \Pr[A(D') \in S] + \delta
\end{equation}

In Equation~\ref{eq10}, $\delta$ is a very small probability value (for example $\delta < 10^{-5}$), and this value expresses that differential privacy violation can occur with a very low probability. For differential privacy to be provided in practice, random noise must be added to the deterministic output obtained from the function to be analyzed. The amount of noise added depends on the ``sensitivity'' of the function. The L1-sensitivity ($\Delta f$) of a function $f: D \rightarrow \mathbb{R}^k$ is defined as the maximum L1 difference in function outputs for any two neighboring datasets $D$ and $D'$ as in Equation~\ref{eq11}:
\begin{equation}
\label{eq11}
\Delta f = \max_{\text{neighboring } D,\, D'} \|f(D) - f(D')\|_1
\end{equation}

Various mechanisms are available for this noise addition process. The Laplace mechanism works by adding noise drawn from a Laplace distribution with mean 0 and scale $\frac{\Delta f}{\varepsilon}$ to the function output $f(D)$ as in Equation~\ref{eq12}~\cite{dwork2006calibrating}:
\begin{equation}
\label{eq12}
A(D) = f(D) + \text{Lap}\!\left(\frac{\Delta f}{\varepsilon}\right)
\end{equation}

The Gaussian mechanism adds noise from $\mathcal{N}(0, \sigma^2)$ distribution to the function; to ensure privacy, Equation~\ref{eq13} must be satisfied~\cite{dwork2014algorithmic}:
\begin{equation}
\label{eq13}
\sigma \geq \frac{\sqrt{2 \ln(1.25/\delta)} \cdot \Delta f}{\varepsilon}
\end{equation}

Differential privacy has some important properties. One of these is post-processing resilience; any deterministic or probabilistic operation applied to a differentially private output does not break the basic privacy guarantee. Another is the composition property; when multiple differentially private algorithms are applied sequentially, the total privacy loss is determined by summing the privacy parameters of individual algorithms as given in Equation~\ref{eq14}~\cite{abowd2018us}:
\begin{equation}
\label{eq14}
\varepsilon_{\text{total}} = \sum_{i=1}^{k} \varepsilon_i
\end{equation}

Additionally, differential privacy also supports group privacy; a change in the data of $t$ individuals simultaneously affects the privacy parameter by $t \cdot \varepsilon$. Differential privacy has been successfully applied in various fields. For example, in the field of public statistics, the United States Census Bureau used differential privacy techniques in publishing the 2020 census data to ensure the protection of individual privacy. Technology companies have also adopted this approach when analyzing user data; for instance, Apple used DP to anonymously analyze user behaviors in iOS operating systems, and Google included similar applications in services like Chrome and Gboard. In the field of machine learning, techniques such as Differentially Private SGD (DP-SGD) based on the principle of adding noise to the SGD algorithm used during model training have been developed.

In conclusion, differential privacy establishes a strong and mathematically provable balance between data analysis and the protection of individual privacy. The robust privacy guarantees it provides make it a fundamental requirement for modern data science applications. With its wider adoption in the future, differential privacy is expected to become widespread as a standard tool in protecting individual privacy.

When these two methods are applied together, both the viewing of data by third parties during the communication process is prevented, and the risk of extracting individual data from model updates is reduced. In this way, while preserving the decentralized data processing advantage of FL, more reliable artificial intelligence training has been performed in areas with high privacy and security requirements, such as health \cite{majmaie2026ssdbfan}.

In this study, a multi-layer artificial neural network (ANN) based model was developed for binary classification problems. The model was created and trained using the PyTorch library. Additionally, differential privacy and homomorphic encryption techniques along with federated learning method, were applied to preserve the model's privacy. The model consists of three fully connected layers. The input layer was dimensioned according to the number of features in the dataset. The first hidden layer has 256 neurons, and the second hidden layer has 128 neurons. Layer normalization and ReLU activation function were applied in both layers. The model's output layer consists of a single neuron and is converted to a probability value with the sigmoid activation function. Overfitting was prevented with the Dropout method (50\%). The model's purpose is to make predictions in the form of binary classification on given data. \cite{al2026zero}.

The datasets used in the study were divided into equal-sized subsets according to the determined number of clients to simulate the decentralized federated learning environment. Each client performed model training independently on its own data subset. Training data for each client was converted to PyTorch TensorDataset format and divided into minibatches via DataLoader (batch size=16).

Model training was performed under the federated learning paradigm. In this method, each client initialized its local model with global model parameters, trained on its own dataset for 5 epochs, then encrypted and sent the updated weights to the central model. The global model summed the homomorphically encrypted weights received from clients, took their averages, and these averages were decrypted to update global model parameters. Subsequently, this updated model is sent back to clients, and clients train their models again for 5 epochs with a learning rate of 0.001 using their local data. Each client training the model for 5 epochs and combining these models into a single model will be referred to as ``Round'' in the later parts of the study.

To provide differential privacy, PrivacyEngine was used during each client's training process. This mechanism protects user data privacy by adding random noise to gradients and calculates the total privacy budget ($\epsilon$, $\delta$). Thus, data leakage was prevented during the federated learning process. Differential privacy parameters sigma, max grad norm, and delta values were chosen as 5.0, 0.5, and $10^{-5}$ respectively, and all training was conducted with these constant values.

At the end of each federated learning round, the global model was evaluated on the test dataset and performance metrics such as accuracy, precision, recall, and F1 score were calculated.

\subsection{Proposed Algorithm}

The developed algorithm comprises two main phases: Preparation and Training, as outlined in Algorithm~\ref{alg:fl-dp-he}. These phases are elaborated below.

\begin{enumerate}
    \item \textbf{Preparation Phase:} In the first phase of the algorithm, the selected dataset (e.g., Framingham heart disease, diabetes, or banking data) is initially divided into training and test subsets. The training data is distributed among $N$ clients in accordance with the FL paradigm, whereby each client receives a unique data partition. The test data is then formatted into a test\_dataloader structure suitable for centralized evaluation. Subsequently, a homomorphic encryption context is established to enable secure processing of model parameters in encrypted form. This mechanism ensures that model weights transmitted from clients to the central server are never exposed in plain text. A central neural network model, referred to as the global model, is then initialized to serve as the foundational architecture of the FL system. Concurrently, differential privacy (DP) parameters are configured. In addition, training hyperparameters are defined. Lastly, for each client, a separate instance of a privacy management mechanism (PrivacyEngine) is instantiated and integrated into the system. These components collectively ensure that the training process is both privacy-preserving and robust.

    \item \textbf{Training Phase:} The FL training process proceeds through a loop that iterates for a predefined number of rounds, denoted as num\_rounds. Each round is structured such that clients contribute to the global model sequentially by submitting their locally trained and encrypted updates. At the beginning of each round, the client\_weights structure (used to collect the encrypted model parameters from all participating clients) is reset to begin a new aggregation cycle.

    \item \textbf{Client Cycle:} In this sub-phase, each client sequentially engages in the training process using its local data. For every client, a dataloader object is created based on the client's assigned dataset. A local copy of the global model, referred to as client\_model, is then instantiated and initialized with the parameters of the current global model to ensure consistency across clients. An optimizer is then constructed for training purposes. Each client's associated PrivacyEngine is invoked from the system to privatize the model, optimizer, and data pipeline. This ensures compliance with differential privacy constraints and provides formal privacy guarantees during local training.

    \item \textbf{Epoch Cycle:} Each client trains its local model over a specified number of epochs. During each epoch, the model is updated using its local data, and the training loss is computed. Once training is done, the trained model's weights are encrypted using homomorphic encryption and appended to the client\_weights structure.

    \item \textbf{Model Aggregation and Evaluation:} Upon the completion of the client cycle, all encrypted models collected in client\_weights are aggregated using the Federated Averaging (FedAvg) algorithm. These aggregated parameters are then decrypted and used to update the global model. This step maintains the core principles of FL by preserving data privacy while ensuring centralized model synchronization. Finally, the updated global model is evaluated using the test\_dataloader. Performance metrics such as accuracy, precision, recall, and F1-score are calculated to assess the model's effectiveness. The training round counter is then incremented, and the cycle continues until the specified number of rounds is completed.
\end{enumerate}

Details are givin in Algorithm~\ref{alg:fl-dp-he} for the proposed approach. Furhermore, in the Preparation phase, the data are split into train/test, training data are partitioned across $N$ clients, and the global model is initialized alongside homomorphic encryption (HE) and differential privacy (DP) settings. In the Training phase, clients train locally and send encrypted updates; the server aggregates them via FedAvg, updates the global model, and evaluates on the test set.

\begin{algorithm}
\caption{Federated learning with differential privacy (DP) and homomorphic encryption (HE)}
\label{alg:fl-dp-he}
\begin{algorithmic}[1]
\State \textbf{Input:} dataset $(X,y)$, number of clients $N$, rounds $T$, local epochs $E$, learning rate $\eta$
\State \textbf{DP params:} noise multiplier $\sigma{=}5$, max\_grad\_norm $=0.5$, $\delta{=}10^{-5}$

\State \textbf{Preparation:}
\State Split $(X_{\text{train}},y_{\text{train}})$ into $N$ client datasets $\{\mathcal{D}_k\}_{k=1}^{N}$; prepare test loader from $(X_{\text{test}},y_{\text{test}})$
\State Initialize HE context $\mathsf{HEctx}$ and global model $w^{0}$
\State Create a DP PrivacyEngine $\mathsf{PE}_k$ for each client $k$

\For{$t=1$ to $T$} \Comment{Global rounds}
    \State Initialize list $\mathcal{W}\leftarrow [\ ]$
    \For{$k=1$ to $N$} \Comment{Client loop}
        \State Initialize client model $w_k \leftarrow w^{t-1}$ and optimizer (Adam, $\eta$)
        \State Make training private using $\mathsf{PE}_k$ (attach DP to model/optimizer/dataloader)
        \For{$e=1$ to $E$} \Comment{Local epochs}
            \State Train $w_k$ on $\mathcal{D}_k$; compute loss
            \State Obtain privacy budget $\epsilon_k$ from $\mathsf{PE}_k$
        \EndFor
        \State Encrypt client parameters $\tilde{w}_k \leftarrow \mathsf{Enc}_{\mathsf{HEctx}}(w_k)$
        \State Append $\tilde{w}_k$ to $\mathcal{W}$
    \EndFor

    \State Aggregate encrypted updates via FedAvg: $\tilde{w} \leftarrow \mathsf{FedAvg}(\mathcal{W})$
    \State Decrypt: $w^{t} \leftarrow \mathsf{Dec}_{\mathsf{HEctx}}(\tilde{w})$
    \State Evaluate $w^{t}$ on the test set; report accuracy, precision, recall, and F1-score
\EndFor
\end{algorithmic}
\end{algorithm}

\section{Experimental Results}

In this study, a federated learning system with enhanced security through homomorphic encryption and differential privacy was tested on different datasets, and the obtained results were analyzed. Experiments were conducted to examine the effect of the number of clients (3, 5, and 10) on training performance. The datasets used are Bank Marketing, Framingham, and Pima Indians Diabetes datasets.

\begin{figure}[htbp]
    \centering
    \includegraphics[width=0.7\linewidth]{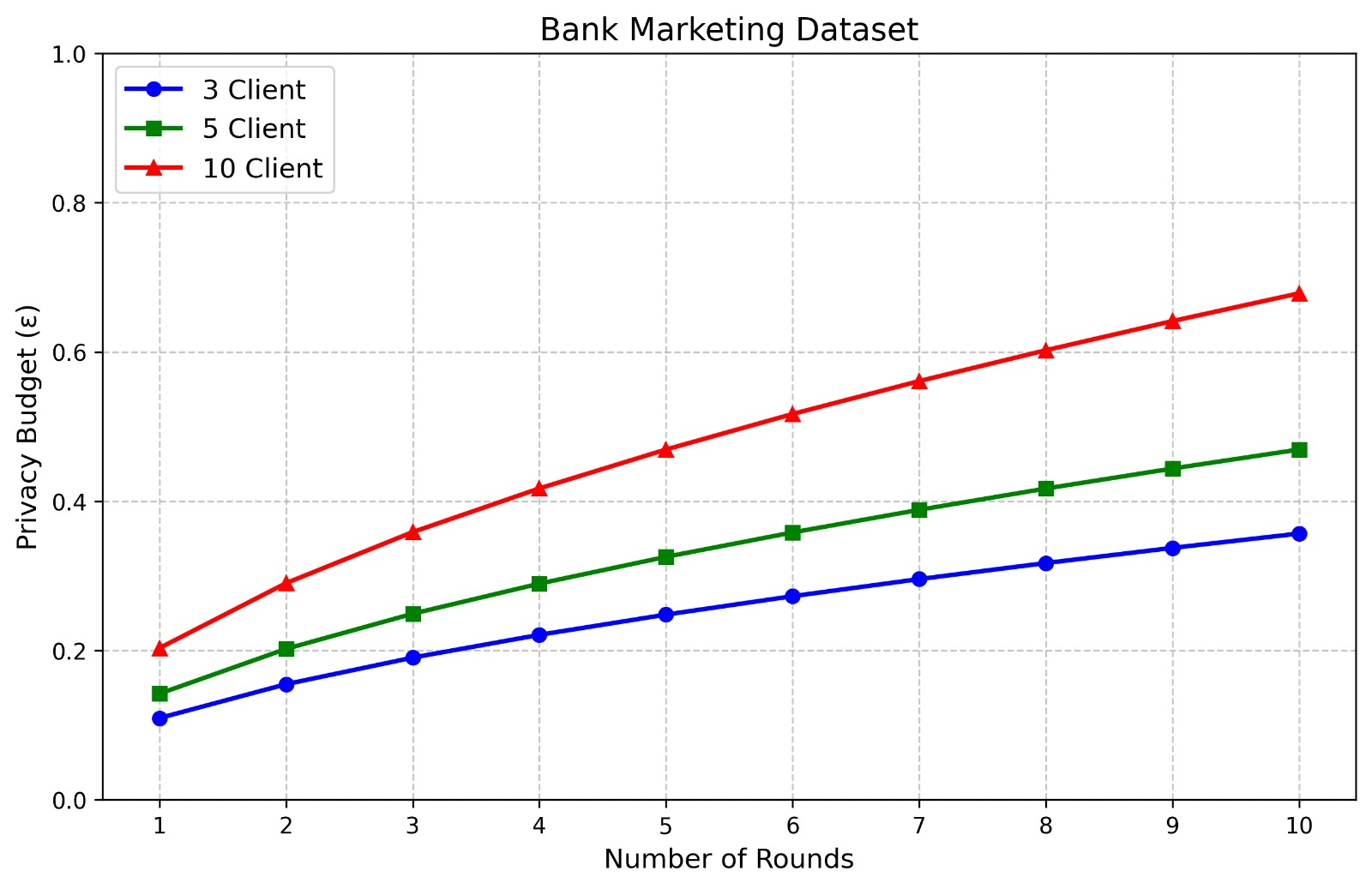}
    \caption{Bank Dataset; Graph showing training results and epsilon values across 10 rounds for the Bank dataset with 3, 5, and 10 clients.}
    \label{bank}
\end{figure}

\begin{figure}[htbp]
    \centering
    \includegraphics[width=0.7\linewidth]{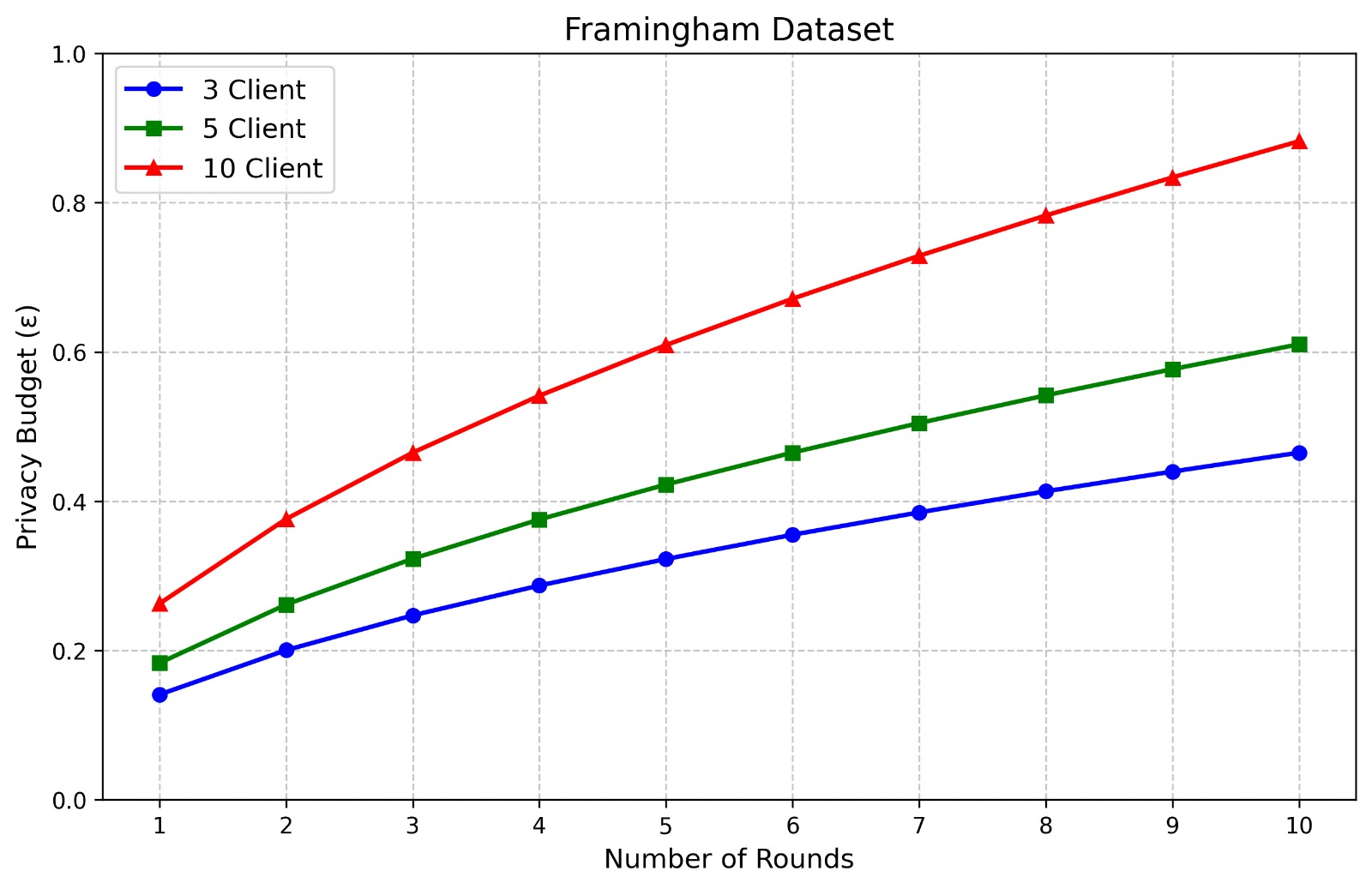}
    \caption{Framingham Dataset; Graph showing training results and epsilon values across 10 rounds for the Framingham dataset with 3, 5, and 10 clients.}
    \label{framingham}
\end{figure}

\begin{figure}[htbp]
    \centering
    \includegraphics[width=0.7\linewidth]{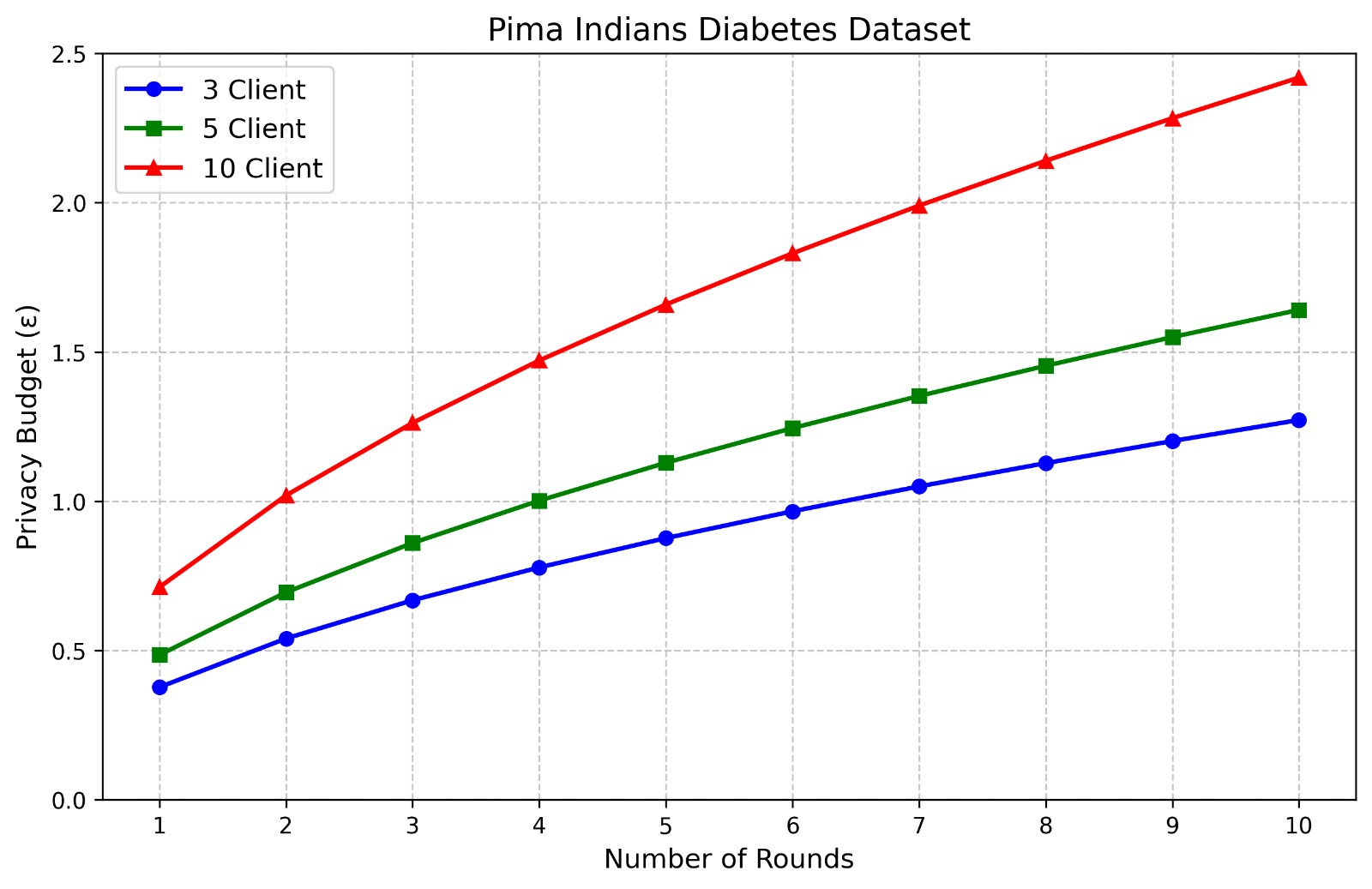}
    \caption{Diabetes Dataset; Graph showing training results and epsilon values across 10 rounds for the Pima Indians Diabetes dataset with 3, 5, and 10 clients.}
    \label{diabetes}
\end{figure}

Figure~\ref{bank}, Figure~\ref{framingham}, and Figure~\ref{diabetes} present the experimental results conducted on 3 different datasets. The training conducted for 10 rounds shows how the dataset is divided into 3, 5, and 10 clients for federated learning implementation and how the privacy level (epsilon value) provided by the system changes during training.

\subsection*{Relationship Between Number of Clients and Privacy Budget ($\varepsilon$)}

In federated learning, as the number of clients increases, the total privacy budget ($\varepsilon$) increases more rapidly. The main reason for this is that the dataset is divided into more parts and each client trains with fewer samples. For example:

In the Bank dataset (10k samples): 3 clients $\rightarrow$ $\varepsilon = 0.3566$ (Round 10); 10 clients $\rightarrow$ $\varepsilon = 0.6785$ (Round 10) (${\sim}2\times$ increase).

In the PID dataset (1k samples): 3 clients $\rightarrow$ $\varepsilon = 1.2719$ (Round 10); 10 clients $\rightarrow$ $\varepsilon = 2.4189$ (Round 10) (${\sim}2\times$ increase).

This situation can be explained by the fact that as the amount of data per client decreases, each local model's update requires more noise. Differential privacy mechanisms add larger noise as the dataset becomes smaller, leading to faster consumption of $\varepsilon$.

\subsection*{Effect of Dataset Size on $\varepsilon$}

In smaller datasets, $\varepsilon$ values reach higher levels. This is particularly related to the decrease in the number of samples per client: PID (1k samples): 10 clients $\rightarrow$ $\varepsilon = 2.4189$ (Round 10); Bank (10k samples): 10 clients $\rightarrow$ $\varepsilon = 0.6785$ (Round 10). This difference stems from the fact that in large datasets, the effect of each individual sample on the model is less significant. In small datasets, since the effect of a single sample on the model is more prominent, the amount of noise added to preserve privacy increases, and $\varepsilon$ grows faster \cite{joshi2026securing}.

\subsection*{Increase in $\varepsilon$ as Training Process Progresses}

With each new round, the cumulative $\varepsilon$ value increases. This results from the composition property of differential privacy: Bank dataset (3 clients): Round~1 $\rightarrow$ $\varepsilon = 0.1094$; Round~10 $\rightarrow$ $\varepsilon = 0.3566$ (${\sim}3.3\times$ increase). PID dataset (10 clients): Round~1 $\rightarrow$ $\varepsilon = 0.7119$; Round~10 $\rightarrow$ $\varepsilon = 2.4189$ (${\sim}3.4\times$ increase). This trend shows that privacy guarantees may weaken in long-term training. In differential privacy, as multiple queries (rounds) are made, the total privacy cost increases; $\varepsilon$ needs to be managed in a controlled manner.

In conclusion: as the number of clients increases, $\varepsilon$ increases faster because each client is trained with less data and the need for noise addition increases. In small datasets, $\varepsilon$ reaches higher levels because the effect of each individual sample on the model is more prominent. As training duration extends (number of rounds increases), $\varepsilon$ increases cumulatively, which may weaken privacy guarantees in the long term.

\begin{figure}[htbp]
    \centering
    \includegraphics[width=0.9\linewidth]{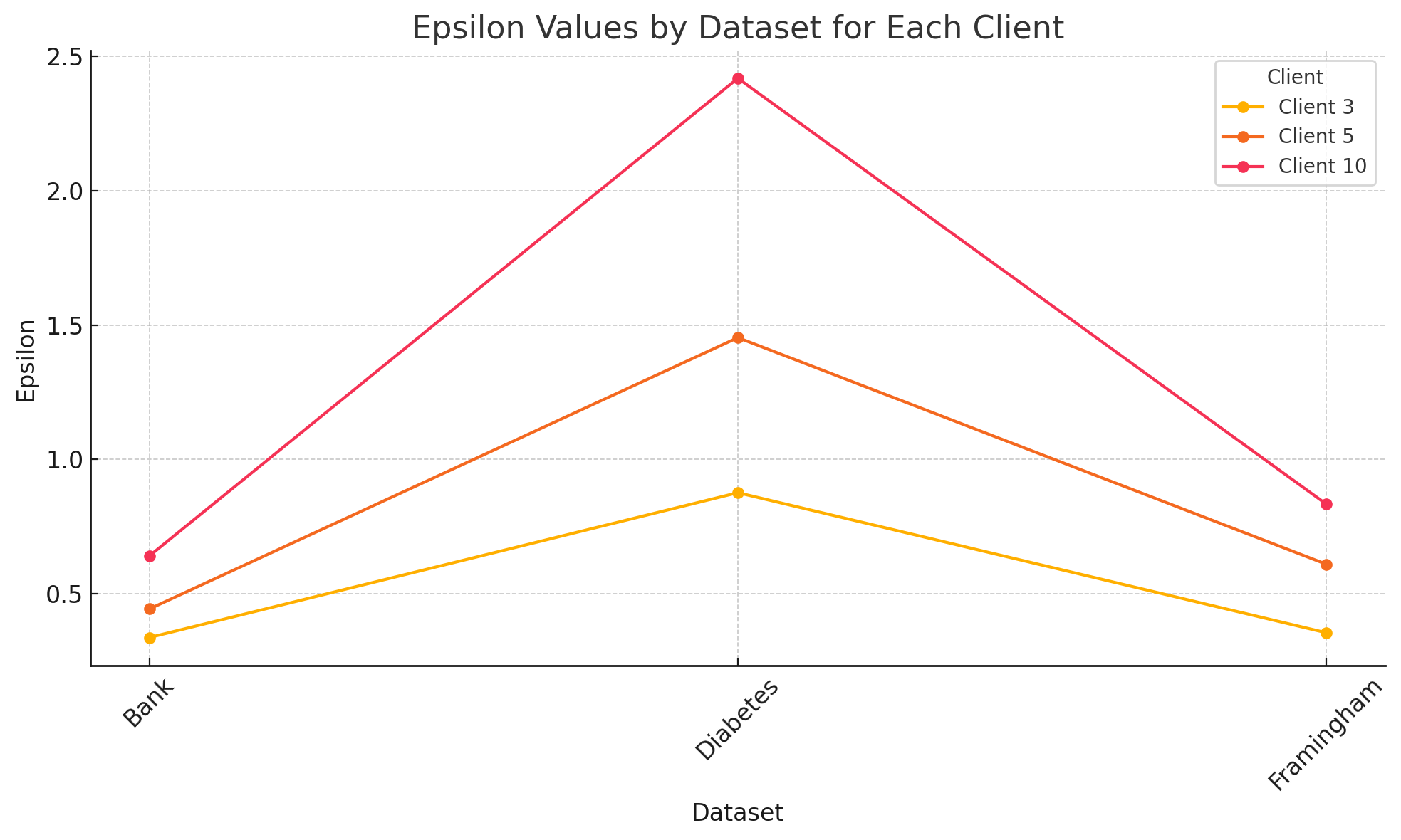}
    \caption{Epsilon results; A graph showing the cumulative epsilon values across training rounds for different numbers of clients and datasets.}
    \label{fig:epsilon}
\end{figure}

Table~\ref{tab:table2}, Table~\ref{tab:table3}, and Figure~\ref{fig:epsilon} present the experimental results obtained from clients tested by dividing three different datasets, namely Framingham, PID, and Bank, each having different data characteristics, into 3, 5, and 10 parts respectively. Accuracy, Precision, Recall, and F1 score were used as evaluation metrics. Additionally, experimental results obtained by applying differential privacy are expressed as ``DP'' and those obtained without application are expressed as ``Without DP''. In the ``DP'' column, in addition to the 4 metrics, there is also an epsilon value, which indicates the privacy level provided by the system at the end of training. The lower the epsilon value, the more private the system is.

\begin{table}
\centering
  \caption{Comparison of method accuracy rates with and without DP.}
  \label{tab:table2}
  \begin{tabular}{lccc}
    \toprule
    Client & Dataset & With DP & Without DP \\
    \midrule
    3  & Bank       & 0.8185 & 0.8620 \\
       & Diabetes   & 0.7200 & 0.7500 \\
       & Framingham & 0.6992 & 0.7147 \\
    \midrule
    5  & Bank       & 0.7836 & 0.8639 \\
       & Diabetes   & 0.7300 & 0.7300 \\
       & Framingham & 0.6496 & 0.7116 \\
    \midrule
    10 & Bank       & 0.8034 & 0.8554 \\
       & Diabetes   & 0.6700 & 0.7100 \\
       & Framingham & 0.6822 & 0.6961 \\
    \bottomrule
  \end{tabular}
\end{table}

In general, the federated learning method supported by privacy and security algorithms has achieved meaningful results in each dataset. It was observed that as the datasets were divided into 3, 5, and 10 clients, the accuracy rates obtained decreased and epsilon increased, meaning that where the data in the system might have come from became more predictable. Additionally, in experiments conducted with the same dataset and number of clients but without applying differential privacy, the model's performance visibly increased, as shown in Tables~\ref{tab:table2} and~\ref{tab:table3}.

\begin{table*}
\centering
  \caption{Comparison of performance metrics with and without differential privacy.}
  \label{tab:table3}
  \begin{tabular}{p{1.5cm}p{2.2cm}ccc ccc}
    \toprule
    Client & Dataset & \multicolumn{3}{c}{With DP} & \multicolumn{3}{c}{Without DP} \\
    \cmidrule(lr){3-5}\cmidrule(lr){6-8}
     & & Precision & Recall & F1-score & Precision & Recall & F1-score \\
    \midrule
    3 & Bank       & 0.8153 & 0.8245 & 0.8199 & 0.8381 & 0.8927 & 0.8646 \\
      & Diabetes   & 0.7568 & 0.5957 & 0.6667 & 0.6964 & 0.8298 & 0.7573 \\
      & Framingham & 0.6744 & 0.7389 & 0.7052 & 0.6738 & 0.8025 & 0.7326 \\
    \midrule
    5 & Bank       & 0.8175 & 0.7572 & 0.7862 & 0.8375 & 0.9011 & 0.8681 \\
      & Diabetes   & 0.7381 & 0.6596 & 0.6966 & 0.6852 & 0.7872 & 0.7327 \\
      & Framingham & 0.6560 & 0.5892 & 0.6208 & 0.6675 & 0.8121 & 0.7328 \\
    \midrule
    10 & Bank       & 0.7844 & 0.8584 & 0.8198 & 0.8028 & 0.9175 & 0.8563 \\
       & Diabetes   & 0.8889 & 0.3404 & 0.4923 & 0.6731 & 0.7447 & 0.7071 \\
       & Framingham & 0.6580 & 0.7229 & 0.6889 & 0.6621 & 0.7675 & 0.7109 \\
    \bottomrule
  \end{tabular}
\end{table*}

The dataset with the highest performance observed was the Bank dataset. Unlike other datasets, all partitioning scenarios produced quite successful results on this dataset. Particularly in the scenario where the data was tested by dividing it into 3 parts, it achieved 81.85\% accuracy, 0.8153 precision, 0.8245 recall, and 0.8199 F1 score, while also providing high-level privacy with a very small epsilon value of 0.3374. This situation shows that the Bank dataset contains much more data compared to other datasets, and the patterns in this dataset have more distinctive features and are less negatively affected by differential privacy processes. Additionally, high recall values highlight the model's reliability, especially in scenarios where false negatives can have critical consequences in financial applications.

Overall, the results show that the proposed privacy-preserving federated learning approach can be successfully applied to various datasets. However, it is also clearly evident that the data distribution among clients directly affects model performance, and in this context, the dataset needs to contain a sufficient amount of data and more advanced weighted model aggregation methods are necessary.

\section{Conclusion}

These results show that data size, number of clients, and training duration directly affect the privacy-efficiency balance in federated learning systems~\cite{xie2024efficiency}. We developed a multilayer artificial neural network for binary classification that integrates differential privacy (DP) and homomorphic encryption (HE) within a federated learning framework. This approach ensured strong data security while maintaining high classification accuracy through local training without centralized data collection, DP noise injection, and secure HE parameter transmission.

Although the distributed architecture provided flexibility and efficient learning, the noise introduced by DP caused a slight performance drop. This highlights the delicate balance required between privacy and accuracy, alongside the communication costs and computational load that affect system scalability.

The findings support the applicability of this privacy-focused model in sensitive domains like healthcare and finance. However, our experiments were limited to a simulated environment with a small number of clients (3, 5, 10) holding relatively large local datasets (e.g., 300--1000 records). In real-world scenarios with numerous clients and sparse local data, performance degradation is anticipated. Future research will focus on exploring different model architectures, advanced aggregation algorithms, optimized DP parameters that inject less noise, and communication optimization techniques to enhance the performance-privacy balance in large-scale applications.

\section{Future Work}

Although the proposed federated learning framework demonstrated promising results in terms of privacy preservation and classification performance, several research directions remain open for future studies. First, the experiments in this study were conducted with a limited number of clients (3, 5, and 10). In real-world federated environments, the number of participating devices can reach thousands or even millions, while the amount of data per client may be significantly smaller. Therefore, future work should investigate the scalability of the proposed framework under large-scale and highly heterogeneous client settings.

Second, communication efficiency remains a critical challenge in federated learning. Future studies may focus on reducing communication overhead through model compression, update sparsification, and adaptive client selection strategies. Such improvements would be essential for deployment in bandwidth-constrained mobile and IoT environments.

Third, the integration of more advanced aggregation algorithms~\cite{onlu2025investigation} may further improve the privacy--utility trade-off. Robust aggregation techniques that are resistant to noisy updates and adversarial behaviors could help mitigate the performance degradation caused by differential privacy noise.

Fourth, exploring alternative deep learning architectures such as transformers, graph neural networks, or hybrid models may enhance performance, particularly for complex and high-dimensional datasets. In addition, adaptive differential privacy mechanisms that dynamically adjust noise levels during training could provide a better balance between privacy and accuracy.

Finally, future research may investigate real-world deployment scenarios and cross-domain applications, particularly in healthcare, finance, and smart city systems, where strict privacy regulations require secure and distributed learning solutions.

\section*{Acknowledgements}
This research did not receive any specific funding.

\subsection*{Disclosure Statement}
The authors declare that they have no conflict of interest.

\subsection*{Author Contributions}
Cagdas Karatas: Conceptualization, Data curation, Formal analysis, Investigation, Methodology, Resources, Software, Validation, Visualization.
Hibanur Karadogan: Conceptualization, Data curation, Formal analysis, Investigation, Methodology, Resources, Software, Validation, Visualization.
Ahmet Yasin Ertug: Conceptualization, Data curation, Formal analysis, Investigation, Methodology, Resources, Software, Validation, Visualization.
Busra Buyuktanir: Supervision, Writing -- original draft, Writing -- review and editing.
Dr.\ Gozde Karatas-Baydogmus: Supervision, Writing -- original draft, Writing -- review and editing.
Dr.\ Kazim Yildiz: Supervision, Writing -- original draft, Writing -- review and editing.

\bibliographystyle{unsrt}  
\bibliography{references}

\end{document}